\documentclass[review,sort&compress]{cas-sc}
\usepackage{amsmath}
\usepackage{graphicx} 
\usepackage[numbers]{natbib}
\usepackage{bm}
\usepackage{hyperref}
\newcommand{\abs}[1]{\lvert#1\rvert}
\usepackage{xcolor}

\begin{document}
\shorttitle{Model of the Current Density in the Electrochemical
  Synthesis of PS Structures with a Lateral Gradient}
\shortauthors{Ospina-Delacruz et al.}
\title[mode=title]{Analytical Model for the Current Density in
the Electrochemical Synthesis of Porous Silicon Structures with a
Lateral Gradient}
\author[1]{C. A. Ospina-Delacruz}[orcid=0000-0002-9609-3031]
\ead{crisalod@gmail.com}
\credit{Investigation; Software; Writing - original draft}
\author[1,2]{V. Agarwal}[orcid=0000-0003-2168-853X]
\fnmark[1]
\ead{vagarwal@uaem.mx}
\credit{Funding acquisition; Investigation; Supervision; Software; Writing - review \& editing}
\author[2]{W. L. Mochán}[type=editor,orcid=0000-0003-0418-5375]
\cormark[1]
\ead{mochan@fis.unam.mx}
\credit{Conceptualization; Funding acquisition; Investigation;
  Methodology; Supervision; Writing - review \& editing}
\address[1]{Centro de
  Investigación en Ingeniería y Ciencias Aplicadas (CIICAp-IICBA),
  Universidad Autónoma del Estado de Morelos (UAEM), Cuernavaca CP
  62209, México}
\address[2]{Instituto de Ciencias F\'isicas, Universidad Nacional
  Autonóma de México, Av. Universidad S/N, Col. Chamilpa, 62210
  Cuernavaca, Morelos, México}
\fntext[fn1]{On sabatical leave at ICF-UNAM from CIICAP-UAEM.}
\cortext[cor1]{Corresponding author}
\begin{keywords}
  Porous Silicon\sep GRIN \sep Reflectance Spectra.\sep
\end{keywords}
\maketitle
\begin{abstract}
  Layered optical devices with a lateral gradient
  can be fabricated through electrochemical synthesis of
  porous silicon (PS) using a position dependent etching current
  density  $\bm j(\bm r_\|)$. Predicting the local
  value of $\bm j(\bm r_\|)$ and the corresponding porosity
  $p(\bm r_\|)$ and etching rate $v(\bm r_\|)$ is desirable for their
  systematic design.
  We develop a simple analytical model for the
  calculation of $\bm j(\bm r_\|)$ within a prism shaped
  cell. Graded single layer PS samples
  were synthesized and their local calibration curves $p$ vs $\bm j$ and $v$
  vs $\bm j$ were obtained from our model and their reflectance spectra.
  The agreement found between the calibration curves from
  different samples shows that from one sample we
  could obtain full calibration curves which may be
  used to predict, design, and fabricate more complex
  non-homogeneous multilayered devices with lateral gradients for
  manifold applications.
\end{abstract}

\section{Introduction}
\label{sec:introduction}
Surfaces with a gradient in the refractive index (GRIN) are of
importance as they allow the engineering and tuning of the optical
phase along their surface, yielding applications such as flat
lenses. Furthermore, there are important biological applications
which combine topographical features with index of refraction
gradients \cite{pierscionek2012gradient}. Preparation of structures
with gradients in other properties have also shown their
usefulness. As an example, the
roughness dependent  response of cells has been tested on
materials used for applications in medical implants
\cite{bai2020bioinspired}.
Gradients have also been employed for several
technological applications such as high-yield screening of catalysts,
sensing materials and bio-molecules \cite{maier2007combinatorial, jayaraman2004construction,
potyrailo2008combinatorial}
Apart from some methods developed by S.E. Fosdick et al. \cite{fosdick2010two},
synthesis of  Ag-Au alloy gradients on steel and chemical composition
gradients of CdS layers
on gold electrodes have also been obtained
\cite{ramaswamy_screening_2011,ramakrishnan_display_2010}.
Among electrochemical methods, changes in the composition and doping density
of  conducting polymers have been used to produce gradients using Indium Tin
Oxide electrodes \cite{inagi2010bipolar}.
Electrochemically induced potential gradients have been shown to catalyze the
reactions and gradient doping of polymers \cite{qin2020bipolar, ishiguro2011gradient}.
As compared to the above mentioned techniques, a relatively economical,
fast and easy approach for the fabrication
of gradient materials has been the use  of asymmetrical
electrode configurations in the electrochemical synthesis of porous silicon \cite{sailor2012porous}.
This technique has the additional advantage of being compatible
with very-large-scale-integration (VLSI) devices that may be
integrated in  microelectronic circuits.

Although porous silicon (PS) initially stimulated the interest of the scientific
community due to its photo-luminescence in the visible range at room
temperature, it has presently
been recognized as a multifaceted optical material due to its large surface
area, bio-compatibility, ease of fabrication and tunable refractive index.
Applications based on porous silicon now cover various fields such as
chemical sensors and biosensors \cite{kumar2020porous,lin_porous_1997,dancil_porous_1999},
microelectronics and micro-mechanical systems (MEMS)
\cite{martinez2016dual}, as well as a
range of optical \cite{estevez2014porous} and opto-electronic applications
\cite{ramadan2020fabrication,namavar_optoelectronic_1993,galkin_mechanisms_2017,gelloz_electroluminescence_2018}. Specifically,
the temporal variation of the current density results in a variation of
porosity along the depth, allowing the easy fabrication of different kinds
\cite{perez2018reflectivity,ariza-flores_white_2012,ghulinyan_porous_2003,girault_porous_2017}
of 1D dielectric multilayered structures.
Although the conventional fabrication of PS yields laterally
homogeneous samples, with invariant structural and optical properties
along the surface, another configuration where a Pt pin electrode is placed
perpendicular to the silicon wafer, produces porous
silicon structures with a lateral gradient in properties such as
thickness, pore size and density, and thus, refractive index and
optical thickness \cite{khung2008using, clements2011mesenchymal}.
The resulting porous surface can have pore sizes ranging from a few nanometers
to  few micrometers  \cite{canham_routes_2018}. The range of the dimensions of the pores and
the corresponding  thickness range on the same chip can be controlled by
adjusting the location of the electrode, the  anodizing current
and the etching time. Different
applications have been found for the resulting structures
\cite{collins2002determining}. For example, Sailor's group
\cite{sailor2012porous, collins2002determining} applied it
for the determination of protein size.
Additionally, a variation in pH of the solution could gate the
trapping/release of bio-molecules at regions with different porosity, a
result potentially useful for drug delivery
applications given the bio-compatible nature of PS.
A similar asymmetric electrode configuration was used to fabricate
multilayered optical filters with lateral a gradient
\cite{li2005painting} and
for developing ethanol sensors \cite{sailor2012porous}.
Additionally, it has been shown that the combination of thermal  oxidation  and
infiltration of TiO$_2$  by atomic layer deposition allows the manufacture
of transparent GRIN
optical elements with a high refractive index contrast \cite{ocier2017tunable}.
Kruger et al. \cite{krueger2018electrochemical} used
a porous silicon 1D GRIN structure infiltrated with polymer to
synthesize flat micro-lens arrays.
Recently, J Wang et al. has shown the fabrication of
a miniature spectrometer with a PS based rugate filter using a
radial interfacial potential distribution \cite{wang2019fabrication}
On the other hand, the effect of different topographical
features present on the same chip, has also been
used to study the dependence of the adhesion of certain cells on the
surface topography \cite{khung2008using}.

For applications such as those mentioned above,
it would be very useful to be able to predict beforehand the spatially dependent
properties of the resulting structures. This would allow the design
and possible optimization of the desired devices.
In this work, we report a simple analytical model to quantitatively
analyze the spatial dependence of the current density during the
electrochemical etching of Si to produce PS nanostructures with a
lateral gradient. A calibration
procedure may then relate the current density to the porosity and the
etching rate. Thus, for a given etching current acting over for a given
time, we may predict the position dependence of the index of
refraction and optical thickness of the resulting inhomogeneous
layer, thus allowing the prediction of the optical properties of
multilayered GRIN structures. The structure of the paper is the
following.
In Section \ref{sec:theory} we develop a
simple analytical model to calculate
the current density at any point over the GRIN structure under
preparation for the case of an electrolytic cell with a simple shape.
In Section \ref{sec:procedures} we provide experimental details for the
synthesis of porous silicon samples with lateral refractive index gradient.
Section \ref{sec:discussion-results} includes experimental results and
their comparison with numerically evaluated parameters. Finally, we
devote Section \ref{sec:conclusions} to our conclusions.

\section{Theory}
\label{sec:theory}
\label{sec:calc-curr-dens}
A gradient in the optical properties of PS systems may be
obtained by electrochemically attacking a Si surface with a position dependent current
density ${\bm j(\bm r)}$, as the porosity and refractive index depend on
the current density, as does the etching rate and the thickness of the
porous layer. Nevertheless, it is challenging to measure the
{\em local} value of $\bm j(\bm r)$ along the surface. As
designing a GRIN structure with predefined parameters such as porosity
and thickness at each point of a GRIN based optical device is desirable
for reproducible device fabrication, we
propose a simple method to compute the current density field at each point
of the device.
Though $\bm j(\bm r)$ may be numerically calculated for any
given arbitrary experimental setup,  here we propose a particularly
simple setup that allows an expression for the current in terms
of a series that is rapidly convergent and each of whose terms is a
simple analytic function. With this setup it is easy to characterize
the dependence of different properties of interest on the current, and
this allows the design and optimization of diverse GRIN PS devices.

\begin{figure}
  \centering
  \includegraphics[width=\textwidth]{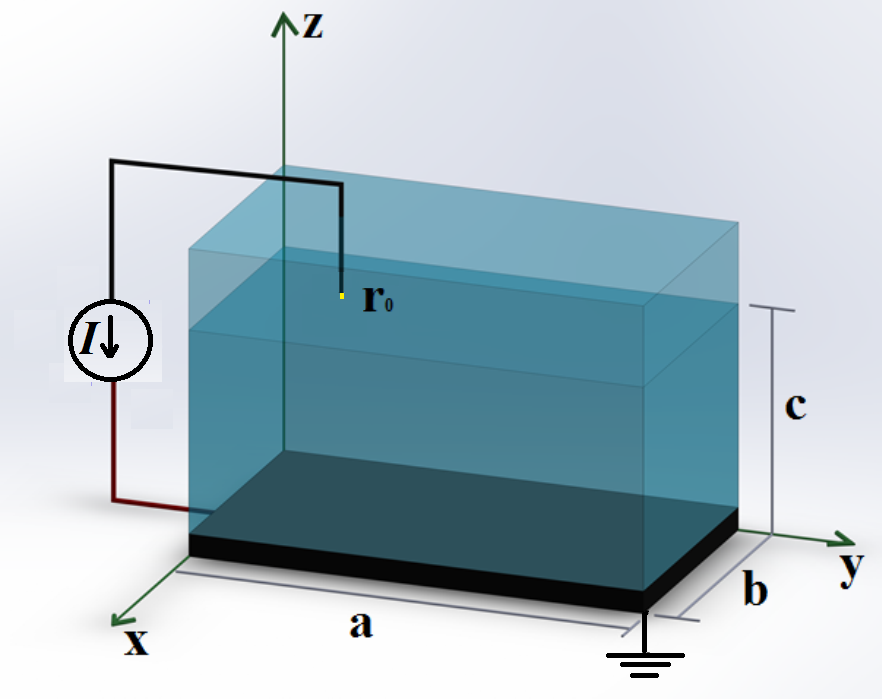}
  \caption{Prism shaped electrolytic cell with a rectangular base of
    sides $a$, $b$ filled with electrolyte up to a height $c$. The
    walls of the cell are insulating and the bottom is completely
    covered by the sample, which is assumed to be a good
    conductor. The tip of the electrode is a thin insulated wire
    with just the tip uncovered and located at $\bm r_0=(x_0,y_0,z_0)$ within the
    liquid. The current flows through the electrolyte from the sample
    to the tip.}
  \label{f:celda}
\end{figure}

We assume that our electrolytic cell has the shape of a rectangular
prism (see Fig. \ref{f:celda}) with a horizontal base of sides $a$ and $b$, and
filled with the electrolyte up to a height $c$. We assume that the
walls of the cell are insulating while the bottom is completely
covered with the sample, which is an
electrically grounded relatively good conductor. Within the electrolyte we position an electrode
which we assume is a thin conducting wire, insulated from the electrolyte
but for a small tip, which we approximate as a
point current source.
Within the electrolyte the current density is
\begin{equation}
  \label{eq:j}
  \bm j=\sigma\bm E,
\end{equation}
with
$\sigma$ the conductivity, $\bm E=-\nabla\phi$ the electric field,
and $\phi$ the electric potential. In a
stationary situation
$\nabla\cdot \bm j = \sigma\nabla\cdot\bm E=0$. Thus, the potential
obeys Laplace's equation within the electrolyte,
\begin{equation}
  \label{eq:laplace}
\nabla^{2} \phi=0.
\end{equation}
Consider now an arbitrary closed surface
$\mathcal S$
within the electrolyte that surrounds completely the tip of the
electrode. Integrating $ \bm j$ over this surface we obtain
\begin{equation}
  \label{eq:I}
\int_{\mathcal S'} d\bm a \cdot \bm j = -I,
\end{equation}
where the prime on $\mathcal S'$ means we remove from the surface a
very small hole through which the wire that feeds the current to the
electrode gets
through, and the sign is consistent with the current direction in
Fig. \ref{f:celda}. Assuming the wire is narrower than any other relevant
distance in the system, we may interpret Eq. \eqref{eq:I} as an
integral over a closed surface of the current \eqref{eq:j} within the
electrolyte, ignoring the actual current within the wire. Thus
\begin{equation}
  \label{eq:gauss}
  \int_{\mathcal S} d\bm a \cdot \bm E = -\frac{I}{\sigma}
\end{equation}
Using Gauss's law, we interpret this equation as a source for the
potential in the form of a point charge
\begin{equation}
  q_0 = -\frac{I}{4 \pi \sigma}
\end{equation}
at the position $\bm r_0=(x_0, y_0, z_0)$ of the
tip of the electrode. Note that $q_0$ would be the total charge,
and it shouldn't be further screened through the permittivity of the
electrolyte.

\begin{figure}
  \centering
  \includegraphics[width=\textwidth]{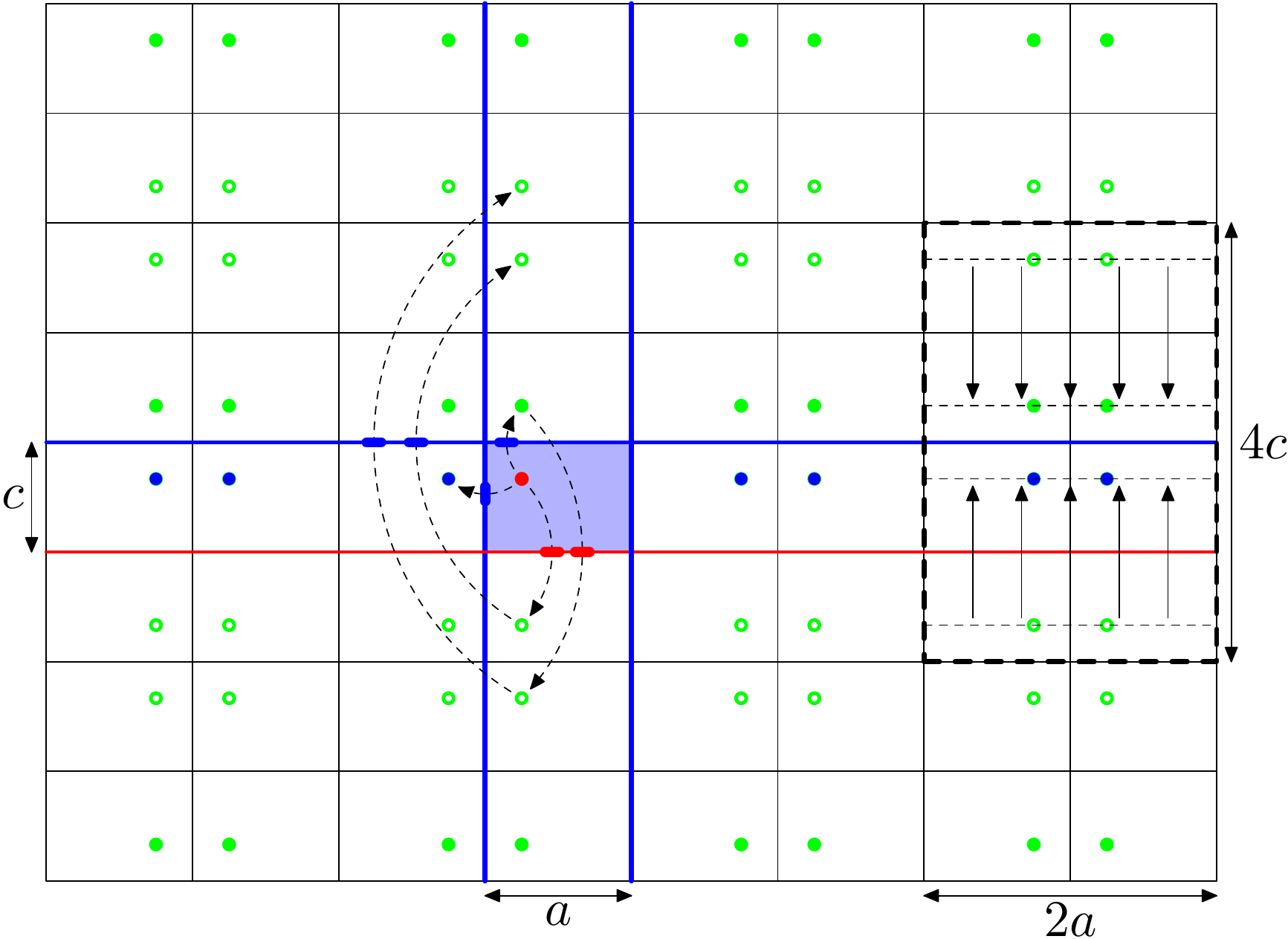}
  \caption{Side view of the $y=y_0$ plane of the fictitious system formed by reflections of
    the electrolytic cell (light blue)
    showing the charge at the tip of the electrode (red dot), its
    images within the plane $z=z_0$ (blue dots), and its images at the
    surface of the sample and the top of the electrolyte (green
    dots). We use solid dots to denote images with the same charge $q_0$
    as that corresponding to the tip of the electrode, and open dots for those of
    opposite charge $-q_0$. The mirror planes that don't invert the charge
    are indicated by blue lines and the mirror plane that inverts the
    charge is indicated by a red line. The relation of some charges and their images is
    indicated by dashed arrows, with a bar to indicate the
    corresponding reflecting wall. The system may be interpreted as a periodic
    lattice with a rectangular unit cell of size $2a\times 4c$, one of
    which is indicated by wide dashed lines, and with a basis of four
    charges $q_0$ and another four charges $-q_0$. Within this cell,
    we indicate its $G=0$ contribution to the electric field (solid
    arrows, see text).}
  \label{fig:sideview}
\end{figure}

The effective conductivity of the relatively thin sample in contact
with the grounded counter-electrode is large enough that we may assume its
surface $z=0$ is an equipotential.
On the other hand, no current can go across the insulating walls of
the cell, situated at $x=0$, $x=a$,, $y=0$ and $y=b$, nor through the
free surface of the liquid at $z=c$. Thus, the problem to solve is Poisson's
equation with a point charge source
\begin{equation}
  \label{eq:poisson}
  \nabla^2\phi(\bm r)=-4\pi q_0\delta^{(3)}(\bm r-\bm r_0),
\end{equation}
with mixed boundary conditions
\begin{equation}
  \label{eq:ground}
  \phi(x,y,0)=0,
\end{equation}
and
\begin{equation}
  \label{eq:E0}
E_x(0, y ,z)= E_x(a, y ,z)=E_y(x, 0 ,z)=E_y(x, b
,z)=E_z(x, y ,c)=0.
\end{equation}
Here, $\delta^{(D)}$ is Dirac's delta function in $D$ dimensions.
This problem may be solved readily using image charge theory \cite{jackson1975classical}.
The potential within the electrolytic cell coincides
within the region $0\le x\le a$, $0\le y \le b$, $0\le z\le c$
with that within an infinite fictitious space  empty but for the point charge
$q_0$ at the tip of the electrode at $\bm r_0$, and an array of its fictitious image
charges situated out of the cell. They include an image on the conducting bottom of the
cell, of the opposite charge $-q_0$ and situated at $(x_0, y_0,-z_0)$,
to guarantee that the bottom stays at potential zero.
There are further images on the four walls of the cell, of the same charge $q_0$
and situated at $(-x_0,y_0,z_0)$, $(x_0,-y_0, z_0)$, $(2a-x_0, y_0,
z_0)$ and $(x_0, 2b-y_0,z_0)$ so that, taken one at a time, no current goes across
the corresponding wall. Similarly, there is an image of the same charge $q_0$ at
$(x_0,y_0,2c-z_0)$ so that no current goes across the
surface of the electrolyte. Furthermore, each image charge has to be
further reflected by each of the aforementioned surfaces, yielding
infinitely many new images so that Eqs. \eqref{eq:ground} and
\eqref{eq:E0} hold when all charges are considered together.
\begin{figure}
  \centering
  \includegraphics[width=\textwidth]{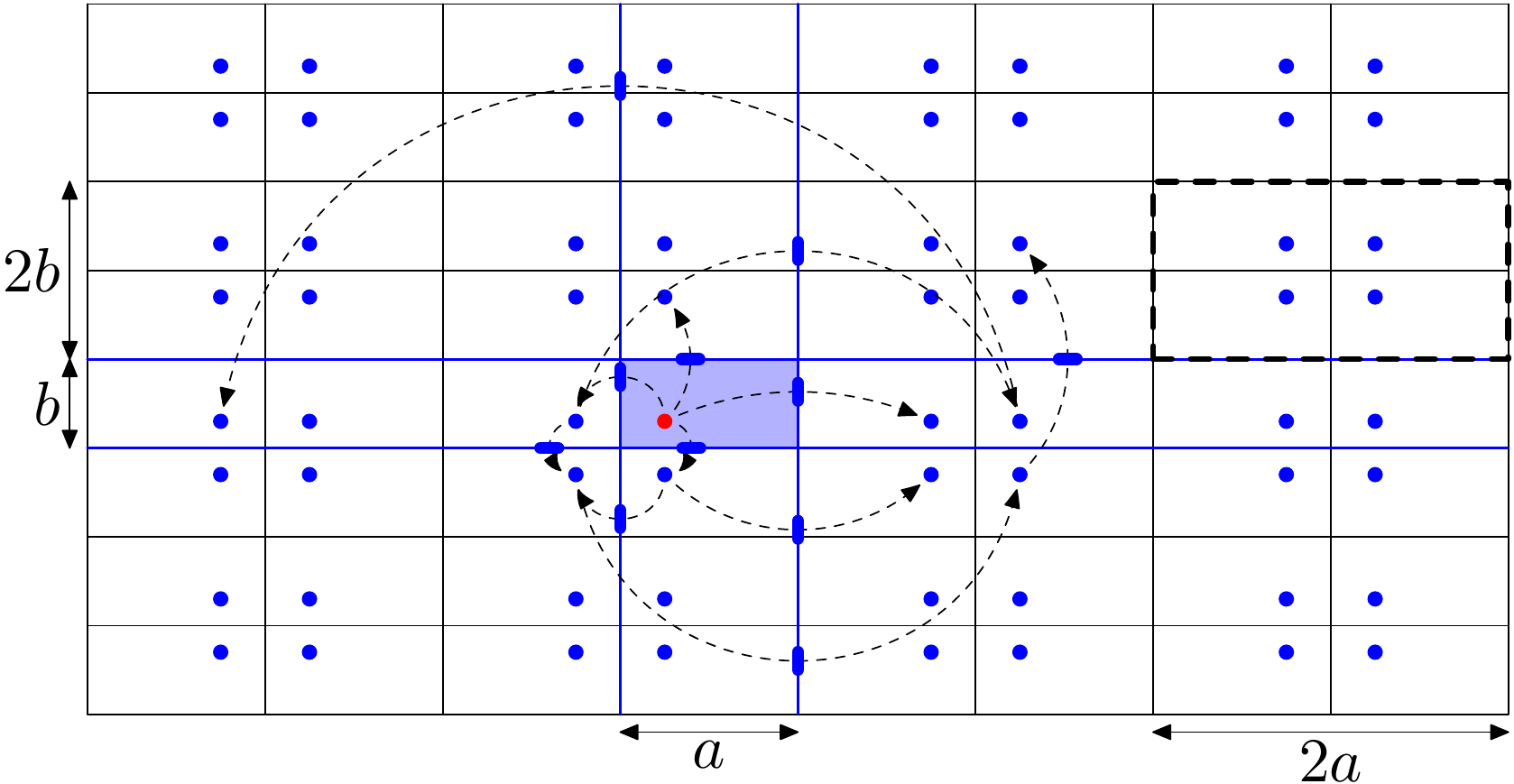}
  \caption{Top view of the $z=z_0$ plane of the fictitious system formed by reflections of
    the electrolytic cell (light blue) showing the charge at the
    tip of the electrode (red dot) and its images on the walls of the
    rectangular cell, the images of its images and so on (blue
    dots). The walls are at $x=0,a$,
    $y=0,b$ (blue lines). The relation of some charges and their images is
    indicated by dashed arrows, with a bar to indicate the
    corresponding reflecting wall. The system may be interpreted as a
    periodic lattice with a rectangular unit cell of size $2a\times 2b$,
    one of which is indicated  by wide dashed lines, and with a basis of four equal
    charges $q_0$ at $(\pm x_0, \pm y_0, z_0)$.}
  \label{fig:topview}
\end{figure}
In Fig. \ref{fig:topview} we illustrate the real charge and all of
its images within the plane $z=z_0$. They form a periodic rectangular
lattice with a unit cell of size $2a\times 2b$ and with a basis of
four equal charges $q_0$ situated at the positions $\bm r_0$, $\bm
r_1=(-x_0, y_0, z_0)$, $\bm r_2=(x_0, -y_0, z_0)$ and $\bm r_3=(-x_0,
-y_0, z_0)$. In Fig. \ref{fig:sideview} we show a lateral view of the
image charges within the plane $y=y_0$.
Each plane of charges upon reflection on the surface of the sample at
the bottom of the cell yields a
plane of image charges of the opposite sign, while each reflection
on the surface of the liquid yields a plane of charges of the same
sign. The side view may be interpreted as a periodic rectangular
lattice with a unit cell of size $2a\times 4c$ containing a basis of eight point
charges, four positive and four negative. A projection onto the $yz$
plane would similarly yield a rectangular lattice with a  $2b\times4c$
unit cell. Putting everything together, the problem is that of solving
Poisson's equation within an orthorhombic lattice of size
$2a\times2b\times4c$ with a basis of 16 point charges of charge $\pm
q_0$ arranged in positive and negative planes.

The electrostatic potential produced by an infinite array of
charges expressed as a sum in real space of Coulomb terms has ill
convergence properties. The potential can be obtained
using the Ewald summation technique
\cite{ewald1921calculation, belhadj1991molecular-dynamics,
  schmidt1997multipole, grzybowski2000ewald, toukmaji2000efficient,
  brodka2004ewald, campione2012ewald}, which yields
two rapidly converging series, one in real space and another one in reciprocal space.
Nevertheless, as we need the potential and the field to obtain the
current density on the surface of the sample on which  no
real nor image charges lie, we may perform a somewhat simpler plane-wise
summation, where we first obtain the potential and field produced by a
simple periodic lattice of charges using a fast convergent sum over
the two dimensional (2D)
reciprocal vectors \cite{nijboer1957calculation, nijboer1958internal,
  de_wette1965internal, watson1981madelung, grindlay1981k},
then sum over the charges that form the basis
of the 2D {\em crystal} plane, and finally sum over all the
planes.

We consider first a system made of identical {\em unit} charges
occupying the positions $\bm R$ of a 2D Bravais lattice $\{\bm R\}$,
which we will take as a rectangular lattice with lattice parameters
$2a$ and $2b$, and lying, for the time being, at the $z=0$ plane.
The potential $\phi$ it produces obeys
\begin{equation}
  \label{eq:poissonRed}
  \nabla^2\phi(\bm r)=-4\pi\sum_{\bm R}\delta^{(2)}(\bm r_\|-\bm
  R)\delta^{(1)}(z)=-\frac{4\pi}{A}\sum_{\bm G}e^{i\bm G\cdot\bm r_\|}\delta^{(1)}(z),
\end{equation}
where $\bm r_\|$ is the projection of the observation position $\bm
r=(x,y,z)$ onto the $xy$ plane, $\{\bm G\}$ is the 2D
reciprocal lattice defined through $e^{i\bm G\cdot\bm R}=1$ for all
$\bm R$ and $\bm G$, $A=4ab$ is the area of the unit cell
and we used the Fourier representation of the 2D
periodically repeated delta function. We introduce a 2D Fourier representation for
the potential
\begin{equation}
  \label{eq:phifourier}
  \phi(\bm r)=\sum_{\bm G}\phi_{\bm G}(z)e^{i\bm G\cdot\bm r_\|}
\end{equation}
where $\phi_{\bm G}(z)$ is the 2D Fourier coefficient of the potential at
the height $z$. Substitution in Eq. \eqref{eq:poissonRed} yields the
ordinary differential equation
 \begin{eqnarray}
 \frac{d^2}{dz^2}\phi_{\bm G}(z)-G^2\phi_G(z)=
   -\frac{4 \pi}{A} \delta^{(1)}(z),
 \end{eqnarray}
for each coefficient, an homogeneous equation for $z\ne0$ which may be trivially
solved. After applying boundary conditions at $z=0$ and regularity
conditions at infinity (for $G\ne 0$), we obtain
\begin{equation}
  \label{eq:phiG}
  \phi_{\bm G}(z)=
  \begin{cases}
    -\dfrac{2\pi}{A}\abs{z}&\text{if }G=0,\\
    \dfrac{2 \pi}{AG}  e^{- G \abs{z}}&\text{if }G\neq 0.
  \end{cases}
\end{equation}
The case $G=0$ corresponds to the potential produced by a uniformly charged plane,
while the case $G\ne 0$ is the potential produced by a sinusoidal
charge density on a plane, given by the
symmetric solution that decays exponentially as we get away, upwards
or downwards, from
the source plane. Substituting into Eq. \eqref{eq:phifourier} yields
\begin{eqnarray}
  \label{eq:phiG1}
 \phi(\bm r) = -\frac{2 \pi}{A}  \abs{z} +{\sum_{\bm G}}' \frac{2 \pi
   }{AG} e^{- G \abs{z}} e^{i \bm G\cdot\bm r_\|},
 \end{eqnarray}
where the prime indicates that the term $G=0$ should be omitted from
the sum.

We consider now the basis of our 2D lattice, with equal charges at the
four positions $(\pm x_0, \pm y_0,0)$. Each of the corresponding four
sub-lattices produces a potential as that in Eq. \eqref{eq:phiG1} but
shifted by the corresponding basis vector, yielding
\begin{equation}
  \label{eq:phiG4}
  \phi(\bm r) = -\frac{8\pi}{A} \abs{z}
    +{\sum_{\bm G}}' \frac{8\pi}{AG} e^{- G \abs{z}}\cos(G_x x_0)
    \cos(G_y y_0) e^{i \bm G\cdot\bm r_\|}.
\end{equation}
Finally, we shift the origin of Eq. \eqref{eq:phiG4} vertically to each of the planes shown in
Fig. \ref{fig:sideview} and multiply by the corresponding charge $q_0$
or $-q_0$ to obtain the contribution to the total potential. The charge $q_0$ corresponds to
the planes with heights of the form $z_0+4nc$ and $-z_0+(4n+2)c$, with
$n=\ldots-2,-1,0,1,2\ldots$ an arbitrary integer, while the charge
$-q_0$ corresponds to the planes with heights $z_0+(4n+2)c$ and
$-z_0+4(n+1)c$.

We concentrate our attention only on the region $0\le z<z_0$ above the
sample but below the electrode. From
Fig. \ref{fig:sideview} we see that the $G=0$
field has contributions only from the charges at $z_0$ and their
images at $-z_0$, and that this contribution is like that of a parallel plate
capacitor $E_{0z}=-4\pi q_0/ab=-16\pi q_0/A$, which corresponds to the
potential $\phi_0=16\pi q_0 z/A$. Other pairs of contiguous planes with opposite charges
contribute to the field elsewhere. The contributions to the potential
from the $G\ne0$ terms are simple geometric series that may be summed
over $n$ analytically and yield the total potential
\begin{equation}
  \label{eq:phitot}
  \begin{split}
    \phi(\bm r)=&\frac{16\pi q_0}{A}\biggl(z+{\sum_{\bm G}}'\frac{2\sinh
      Gc}{G \sinh 2Gc}\cosh G(c-z_0)\cos G_x x_0 \cos G_y y_0\\
    &\times \sinh Gz\, e^{i\bm G\cdot\bm r_\|}\biggr).
  \end{split}
\end{equation}

Notice that the terms of the sum above converge exponentially to zero
as $G$ increases for any $z$ such that $0\le z <z_0 \le c$, and thus
the sum is convergent. The divergence in the limit $ z
\longrightarrow z_0 $ corresponds to the singularity of the potential
at the position of a point charge, and it is not worrisome, as we are
interested in the field close to bottom $z=0$ of the cell.

The current density normal to the surface may now be obtained as
$j_\perp=-\sigma\partial\phi/\partial z$,
\begin{equation}
  \label{Eq:J}
  \begin{split}
    j_\perp(\bm r) =& \frac{I}{ab}\biggl(1+{\sum_{\bm G}}'\frac{2\sinh
      Gc}{\sinh 2Gc}\cosh G(c-z_0)\cos G_x x_0 \cos G_y y_0\\
    &\times \cosh Gz\, e^{i\bm G\cdot\bm r_\|}\biggr).
  \end{split}
\end{equation}
Eq. \eqref{Eq:J} is a very rapidly convergent expression that allows,
in particular, to calculate the current $j_\perp(x,y,0)$ that attacks
the substrate to produce the PS sample. As expected, the current
at the surface $z=0$ of the sample becomes more homogeneous as the
distance $z_0$ to the tip of the electrode and the height of the
electrolyte $c>z_0$ increase, as each term in the sum decays
approximately exponentially with increasing $z_0$.
Integrating over the surface of the sample we verify
\begin{equation}
  \label{eq:intj}
  \int_0^adx\int_0^bdy\,j_\perp(\bm r)=I,
\end{equation}
as the contributions of opposite non-null reciprocal vectors cancel
out: as expected, all the charges that leave the sample find
their way  to the cathode across the electrolyte.

Using Eq. \eqref{Eq:J} we can correlate the local current density at
the sample as it is prepared with its geometric, structural and optical
properties.

\section{Procedures}
\label{sec:procedures}
\begin{figure}
  \centering
  \includegraphics[width=.8\textwidth]{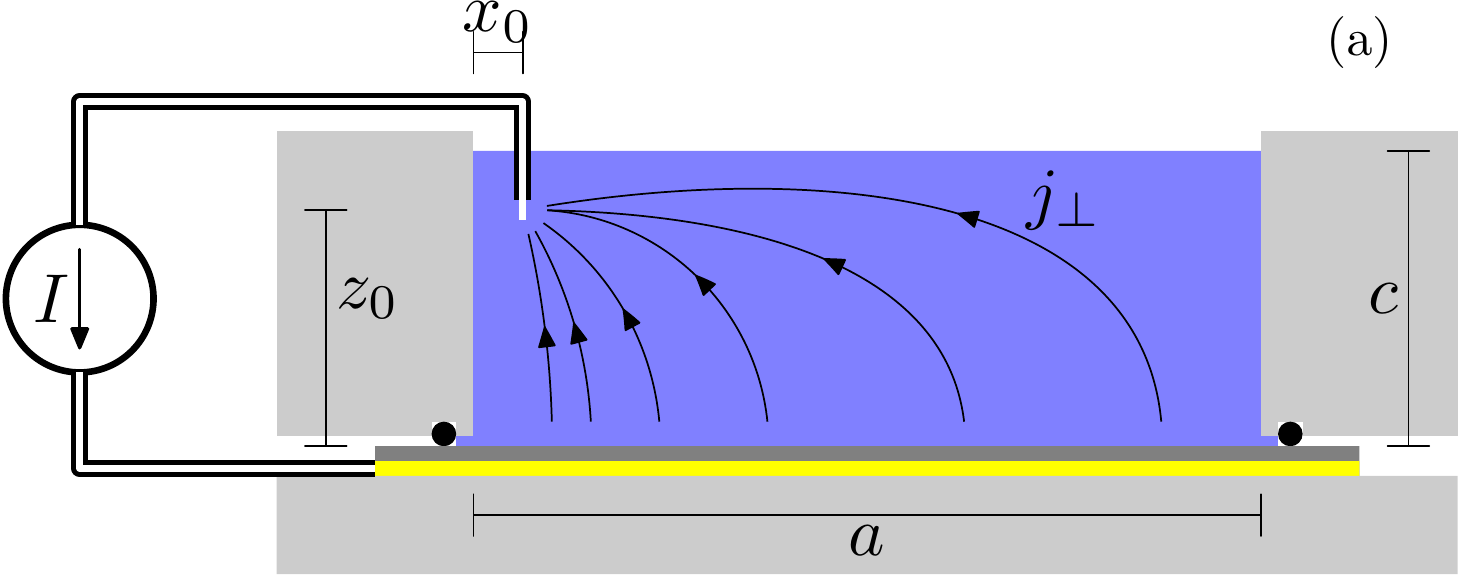}\\[12pt]
  \includegraphics[width=.8\textwidth]{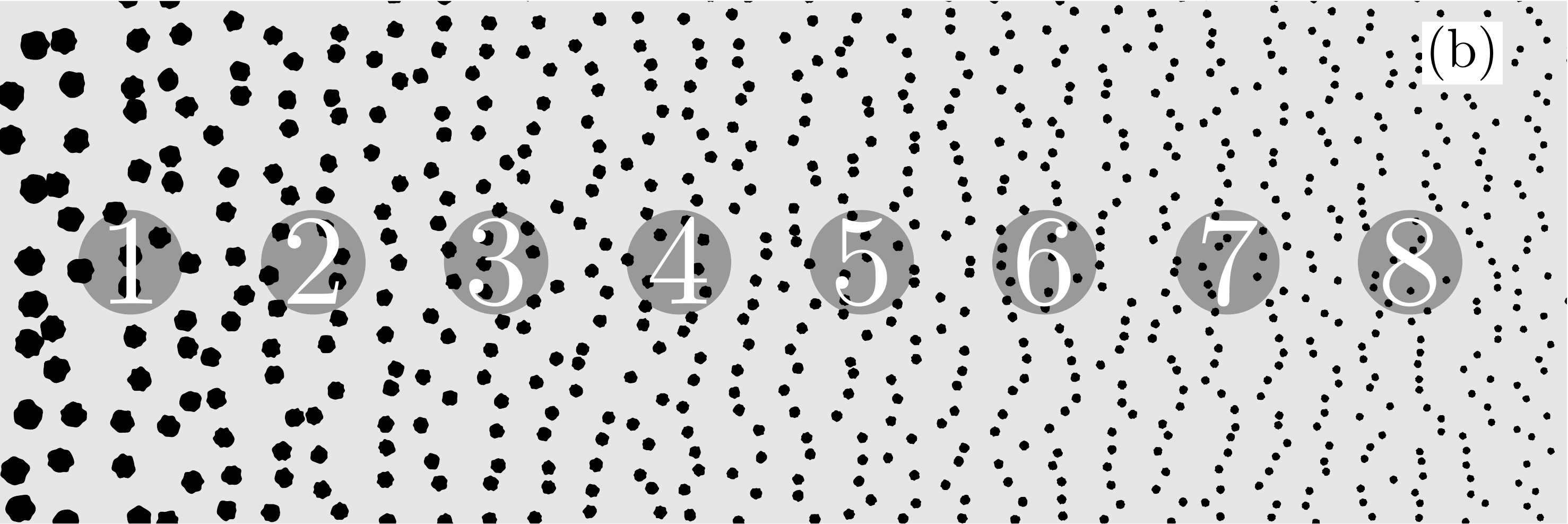}\\[12pt]
  \includegraphics[width=.8\textwidth]{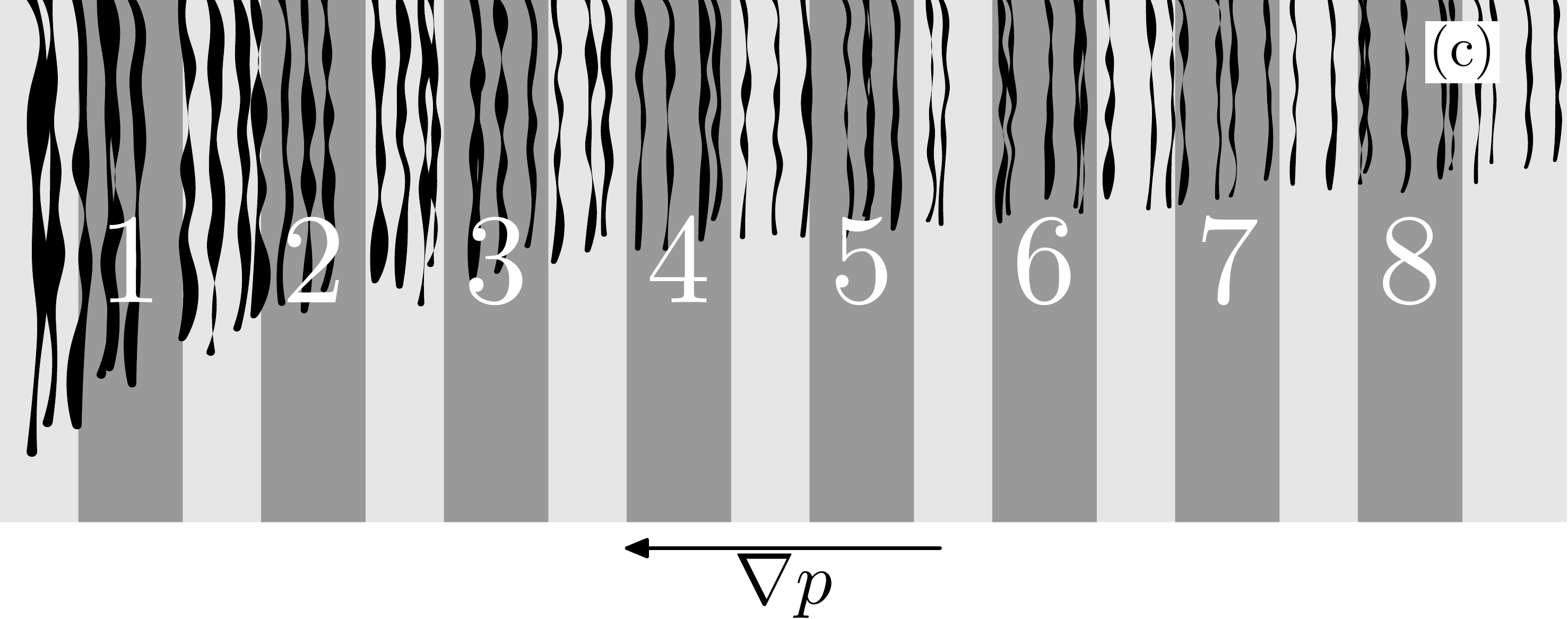}

  \caption{(a): Schematic experimental set-up for
    the fabrication of GRIN PS structures. We show a side view of the
    electrochemical cell with the shape of a
    rectangular prism. We indicate the Si wafer (dark
    gray), the Teflon cell (light gray), the sealing o-ring (black
    circles), the insulated feeding cable (white lines within black
    lines), the uncovered platinum tip (white) within the
    electrolyte (blue), the electrical contact below the sample
    (yellow) and the controlled current source. We indicate some geometrical
    parameters of the arrangement $x_0$, $z_0$, $a$ and $c$
    and a few schematic current flux
    lines. (b) and (c): Top and side schematic views of expected resulting sample
    illustrating eight regions $1\ldots8$ of decreasing porosity and thickness,
    illustrating the changes in pore radii,
    length, and density and the porosity gradient.}
  \label{fig:DE1}
\end{figure}

In Fig. \ref{fig:DE1} we show a schematic representation of our
experimental setup to manufacture GRIN PS structures.
As discussed above, we fabricated an electrolytic cell with the shape of a
rectangular prism on the bottom of which we place a Si wafer which makes
electrical contact to a brass plate
which together make the anode. The
cathode consists of a platinum wire insulated except for a small region
at its tip, which we take as a point current drain.
We also show a schematic drawing of the expected GRIN PS structures,
with a gradient in the porosity, and in the thickness, density and depth
of the resulting pores.

\subsection{Fabrication of PS  GRIN Single Layers}
\label{sec:fabrication-ps-grin}
Single layer PS structures were fabricated through electrochemical
anodization of (100) oriented $p$-type crystalline Si wafer
(resistivity 0.002 - 0.005 $ \Omega\,\text{cm}$), under galvanostatic
conditions. The process was performed at room temperature with an
electrolytic mixture of aqueous hydrofluoric
acid (HF) (concentration 48\% wt), glycerol (purity 99.8 \% wt) and
ethanol (purity 99.9\% wt) in a 3:7:1 volume ratio.
After the anodizing process, the samples were rinsed with
ethanol (purity: 99.9 $ \% $  wt). The electrolytic cell had the shape
of a rectangular prism as shown in Fig. \ref{fig:DE1}, with a base of
sides $a=2.01\,\text{cm}$ and $b=1.48\,\text{cm}$, with an uncovered sample area
of about $3\,\text{cm}^2$. The height of the liquid was set at
$c=1\,\text{cm}$. The cathode was a
platinum wire with a diameter of $0.41\,\text{mm}$ insulated with
Teflon tape, with an uncovered
tip located at $\bm r_0=(x_0,
y_0,z_0)=(0.2\,\text{cm},0.74\,\text{cm},0.9\,\text{cm})$.
We manufactured three samples $S_1$, $S_2$ and $S_3$ by applying a
current $I_1=5\,\text{mA}$, $I_2=10\,\text{mA}$ and
$I_3=20\,\text{mA}$, respectively, during a time of $t=250s$.

\subsection{Reflectance Measurement}
\label{sec:refl-meas}
We measured the absolute reflectance at different positions over our
samples using Universal reflectance accessory  equipped {\em Perkin Elmer Lambda 950}
ultraviolet-visible-near infrared (UV-Vis-NIR) spectro-photometer, with a rectangular spot size of
$0.7\text{mm}\times2.0\text{mm}$, a slit width of $0.6\text{mm}$ in
a wavelength range from $300\text{nm}$ to
$1400\text{nm}$.

For interpretation of the data we used
the standard formula \cite{stenzel2015physics} for the reflectance of
a three media system:
an air environment  (0), the PS film (1) and the crystalline Si (c-Si) substrate (2),
\begin{equation}\label{Eq:ECMR}
  R=\left |\frac{r _ {_ {01}}r _ {_ {12}} e^{2i\psi}}{1+r _ {_
        {01}}r _ {_ {12}} e^{2i\psi}}\right |^2
\end{equation}
where $r_{01}$ and $r_{12}$ are the Fresnel reflectance
coefficients of the air/PS and PS/c-Si interfaces \cite{sattler2017silicon}
and $\psi=k_1^\perp d$, where $k_1^\perp$ is
the component of the wave-vector perpendicular to the surface within
the porous layer of thickness $d$. The reflectance depends on the
thickness of the film and on the index of refraction $n_1=n_{\text{PS}}$ of the porous
layer, which we relate to the porosity through Bruggeman's
effective medium theory \cite{theiss1997optical},
\begin{equation}\label{Eq:Brugg}
 p\frac{1-\epsilon_1} {1+\epsilon_1} + (1-p)
  \frac{\epsilon_2 - \epsilon_1}
  {\epsilon_2+\epsilon_1}=0,
\end{equation}
where $\epsilon_2=\epsilon_{\text{Si}}$ is the dielectric function of
the c-Si substrate and
$\epsilon_1=n_{\text{PS}}^2$ is the dielectric function of the PS layer.
We fitted the reflectance spectra using the porosity $p$ and the
thickness $d$ and a scale calibration factor $s$ as fitting
parameters, and thus, we obtained the porosity
$p$ and the etching rate $v=d/t$ for different values of the local
current density $j_\perp$ expected at several positions on different
samples.

\subsection{Morphological studies}

The morphology of the etched porous layers (plan-view and
cross-sectional) was observed with a {\em Hitachi SU1510} scanning electron microscope
(SEM).


\section{Results and Discussion}
\label{sec:discussion-results}
\subsection{Current Density}
\label{sec:current-density}
\begin{figure}
  \centering
  \begin{picture}(400,300)
    \put(0,0){\includegraphics[width=.8\textwidth]{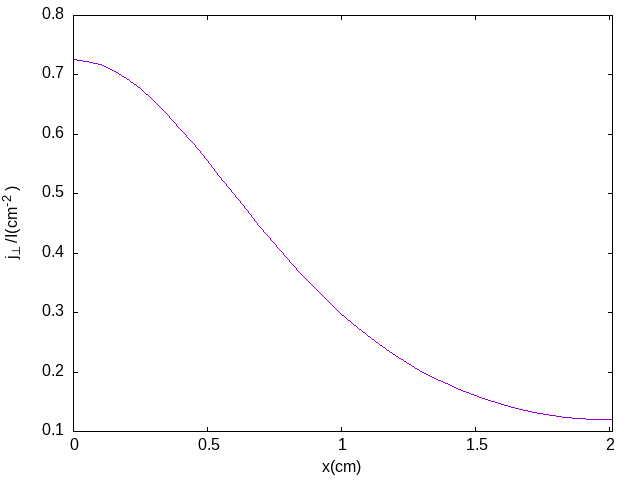}}
    \put(160,140){\includegraphics[trim=50 50 70 80,clip,width=.4\textwidth]{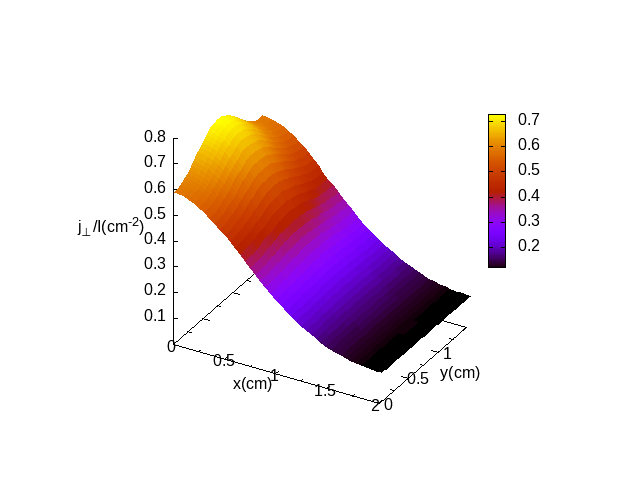}}
  \end{picture}
  \caption{Normalized current density $j_\perp(x,y,0)/I$ for a system
    as in Figs. \ref{f:celda} and \ref{fig:DE1} for a cell of
    length $a = 2.01\,\text{cm}$ and width $b=1.48\,\text{cm}$, with
    the surface of the electrolyte at a height $c=1.0\,\text{cm}$, with a
    point-like electrode at $\bm r_0=(0.2\,\text{cm},
    0.74\,\text{cm},0.9\,\text{cm})$,  calculated through
    Eq. (\ref{Eq:J}) along the center line
    $y=0.74\,\text{cm}$. Inset: current over the whole surface.}
  \label{fig:DR1}
\end{figure}
In Fig. \ref{fig:DR1} we show the normalized current density $j_\perp(x,y,0)/I$ calculated through
Eq. \eqref{Eq:J} at the surface of the sample for our cell, as shown
in Figs. \ref{f:celda} and \ref{fig:DE1} and described in
Sect. \ref{sec:fabrication-ps-grin}). For these parameters we obtained
a large range of values for $ j_\perp/I\sim 0.11-.72\,\text{cm}^{-2}$
that vary mostly along the $x$ direction. Of course, this would differ
for cells with different aspect ratios and for different positions of
the electrode.

\subsection{Samples }
\label{sec:ps-grin-single}
\begin{figure}
 \centering
 \includegraphics[width=\textwidth]{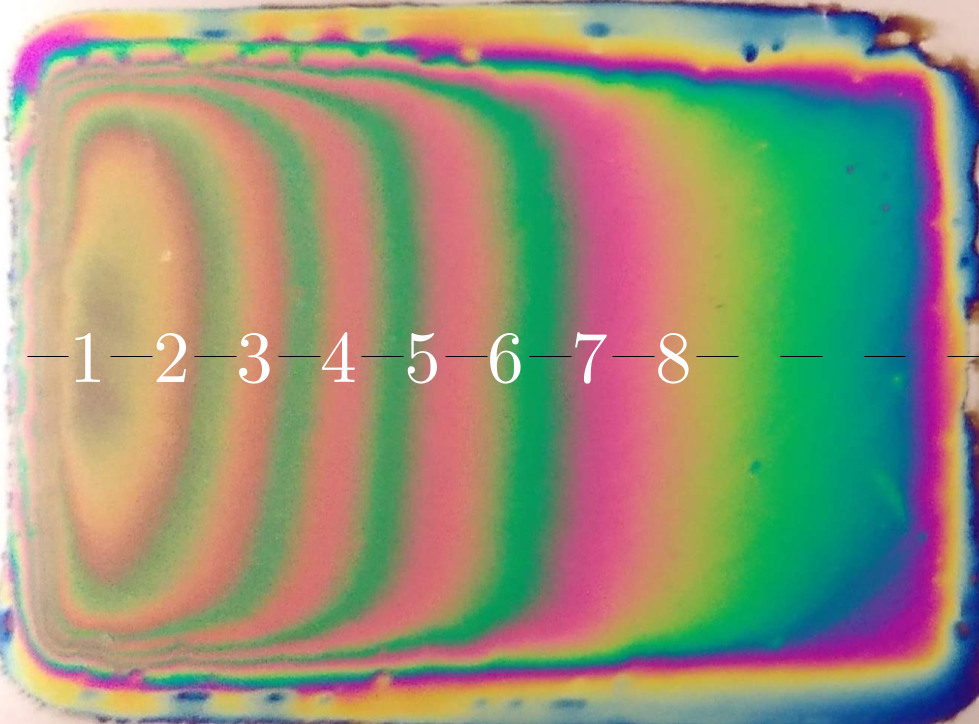}
  \caption{ Photograph of sample $S_3$ prepared as described in
    Sec. \ref{sec:fabrication-ps-grin} using the same parameters as
    in Fig. \ref{fig:DR1} with a current $ I_3=20 \text{mA}$,
    during a time $t=250\text{s}$. We indicate the center-line with a
    dashed line and eight regions $x_i$
    $(i=1, 2,...8)$ centered at positions,
    $x'_1=0.2\text{cm}$,  $x'_2=0.4\text{cm}$, $x'_3=0.6\text{cm}$,
    $x'_4=0.8\text{cm}$, $x'_5=1\text{cm}$, $x'_6=1.2\text{cm}$,
    $x'_7=1.4\text{cm}$, $x'_8=1.6\text{cm}$, with $y_n=0.74\,\text{cm}$
    on which we measured reflectance spectra. Here,
    $x'_n=x_n+0.14\,\text{cm}$ is the position from the left side sealing o-ring.}
  \label{fig:SI3}
\end{figure}

In Fig. \ref{fig:SI3} we show a photograph of one of our samples,
$S_3$,
prepared as described in Sec. \ref{sec:fabrication-ps-grin}, with the
same conditions as those corresponding to Fig.~\ref{fig:DR1} applying
a current
$I_3=20\,\text{mA}$ during a time $t=250\,s$. We notice a series of
visible interference fringes that get wider as we move towards the
right end of the sample, consistent with a inhomogeneous film that
gets thinner towards the right, and qualitatively consistent with
Fig. \ref{fig:DR1} that shows a current density that decays towards
the right. Equal color lines seem to correspond to iso-current lines of
Fig. \ref{fig:SI3}, except very close to the the walls of the cell,
where the film rapidly becomes very narrow. The reason is that there
is a small region below the cell walls where the electrolyte penetrates
up to the sealing o-ring (see upper panel of
Fig. \ref{fig:DE1}) and towards which the current leaks from the
neighborhood of the wall. The width of this region is
$w=0.14\,\text{cm}$. Thus, we may expect our formula \eqref{Eq:J}
to {\em fail} within that region and close to the border, at distances of the order of the radius
of the cross-section of the o-ring, about a millimeter. The
figure shows eight positions
$x'_n=0.2\,\text{cm},0.4\,\text{cm},\ldots,1.6\,\text{cm}$ on which we
measured the near normal incidence reflection spectra of the
sample. Here, we defined $x'_n=x_n+w$ as the distance to the edge of
the etched sample, which corresponds to the position of the sealing o-ring,
whereas $x_n$ is the corresponding distance to the wall of the
cell.

From Eq. \eqref{Eq:J} we evaluate the
normalized current density $j_\perp/I$ expected at the positions $n$ indicated in
Fig. \ref{fig:SI3}. We included an estimated uncertainty $\Delta
  x_n=\Delta x$ in the
distances $x_n$ to the wall of the cell, and the corresponding (asymmetrical)
uncertainty in the computed current density. This
uncertainty arises from the uncontrolled deformation of the o-ring as
it is pressed against the wafer to seal in the electrolyte, so that
its distance $x'_n-x_n\approx 0.14\text{cm}\pm\Delta x$ to the wall of
the cell is not well determined. We estimate $\Delta x$
as the radius of the o-ring (Table \ref{t:xj}).
\begin{table}
  \centering
  \renewcommand{\arraystretch}{1.1}
    \begin{tabular}{rrr}
     $x'_n$(cm) &$x_n$(cm)&$j_\perp/I$ ($\text{cm}^{-2})$\\
     from edge& from wall&\\
     \hline
     0.2 &$ 0.06\pm0.05$ & $0.722^{+0.003}_{-0.007}$\\
     0.4 &$ 0.26\pm0.05$ & $0.672^{+0.018}_{-0.021}$\\
     0.6 &$ 0.46\pm0.05$ & $0.577^{+0.027}_{-0.028}$\\
     0.8 &$ 0.66\pm0.05$ & $0.464^{+0.028}_{-0.028}$\\
     1.0 &$ 0.86\pm0.05$ & $0.360^{+0.025}_{-0.023}$\\
     1.2 &$ 1.06\pm0.05$ & $0.275^{+0.019}_{-0.018}$\\
     1.4 &$ 1.26\pm0.05$ & $0.211^{+0.014}_{-0.013}$\\
     1.6 &$ 1.46\pm0.05$ & $0.167^{+0.010}_{-0.009}$\\
    \end{tabular}
  \caption{Normalized current density $j_\perp/I$ expected at different positions $x'_n$.}
  \label{t:xj}
\end{table}

\begin{figure}
  \centering
  \includegraphics[width=.3\textwidth]{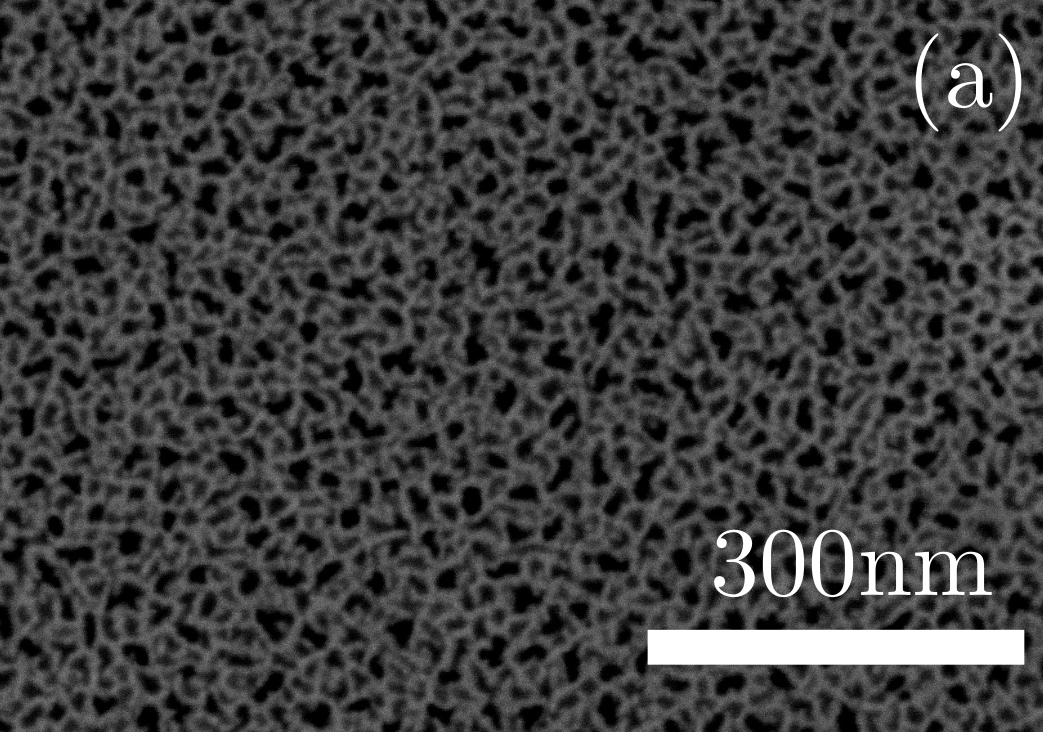}
  \includegraphics[width=.3\textwidth]{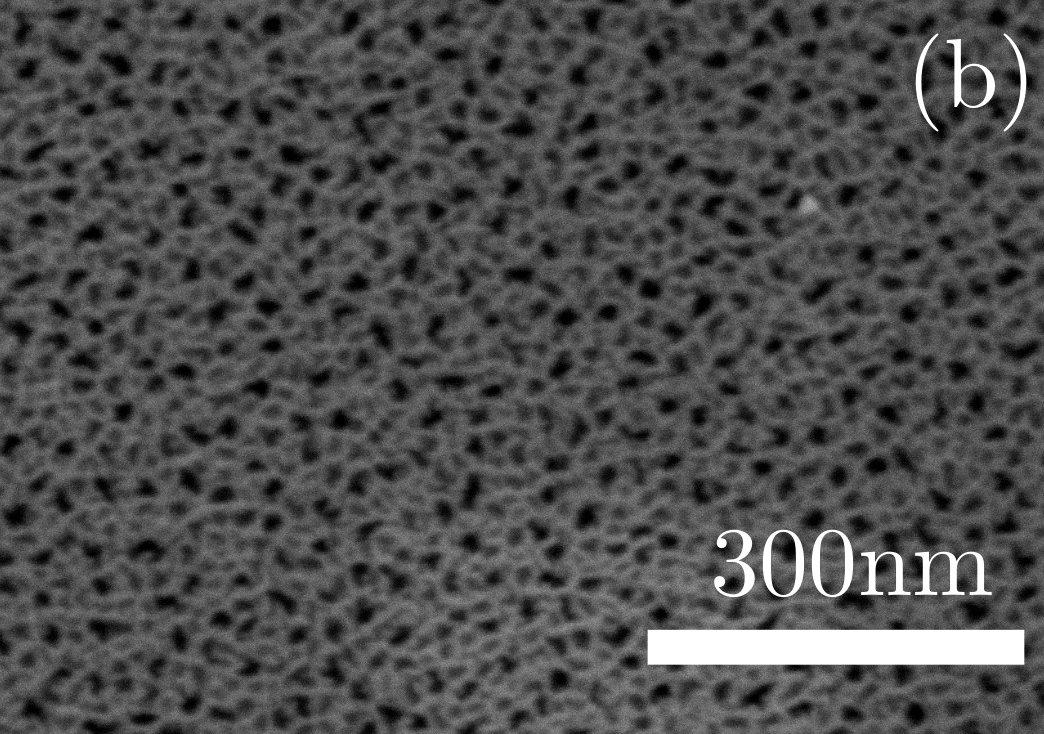}
  \includegraphics[width=.3\textwidth]{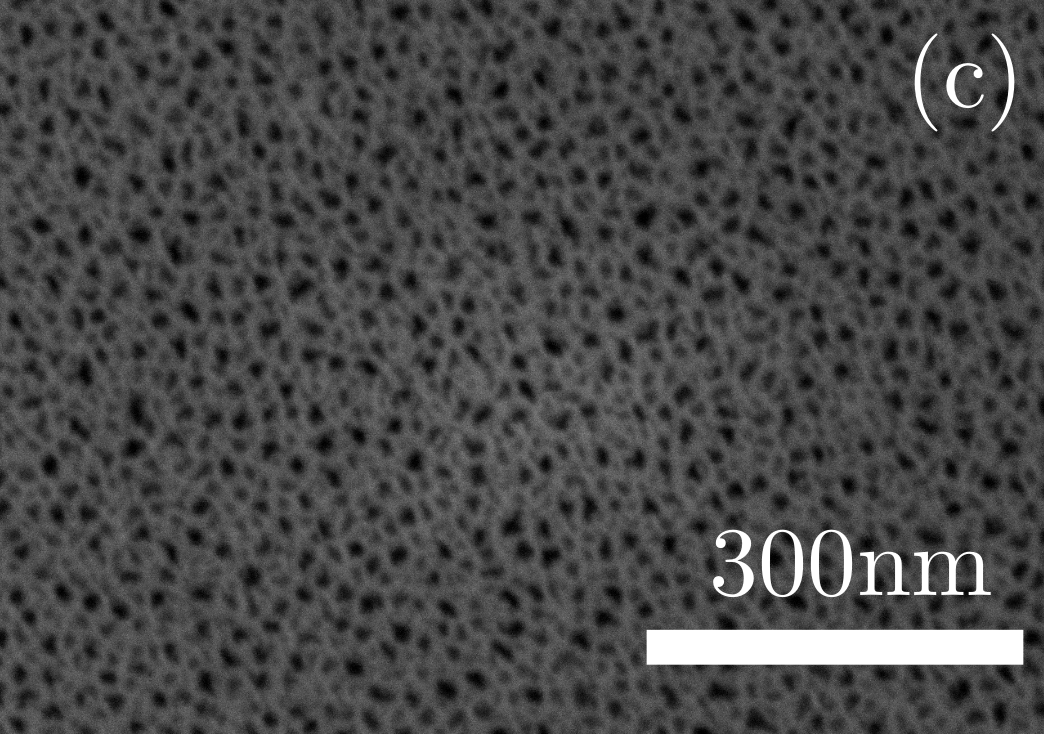}
  \includegraphics[width=.3\textwidth]{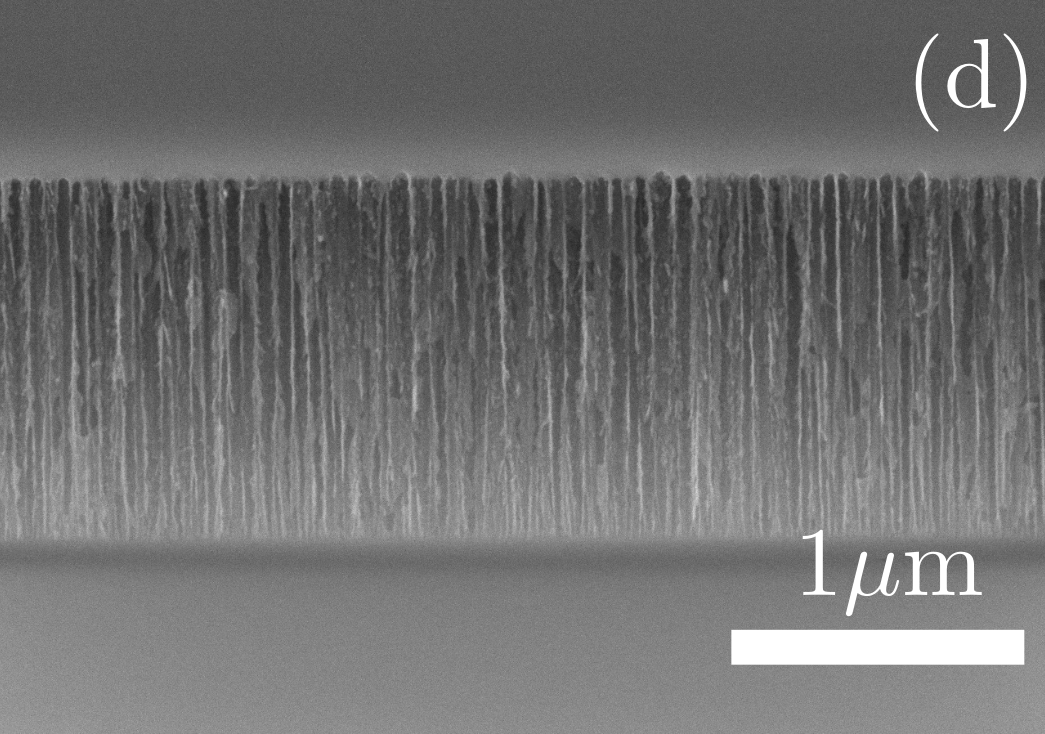}
  \includegraphics[width=.3\textwidth]{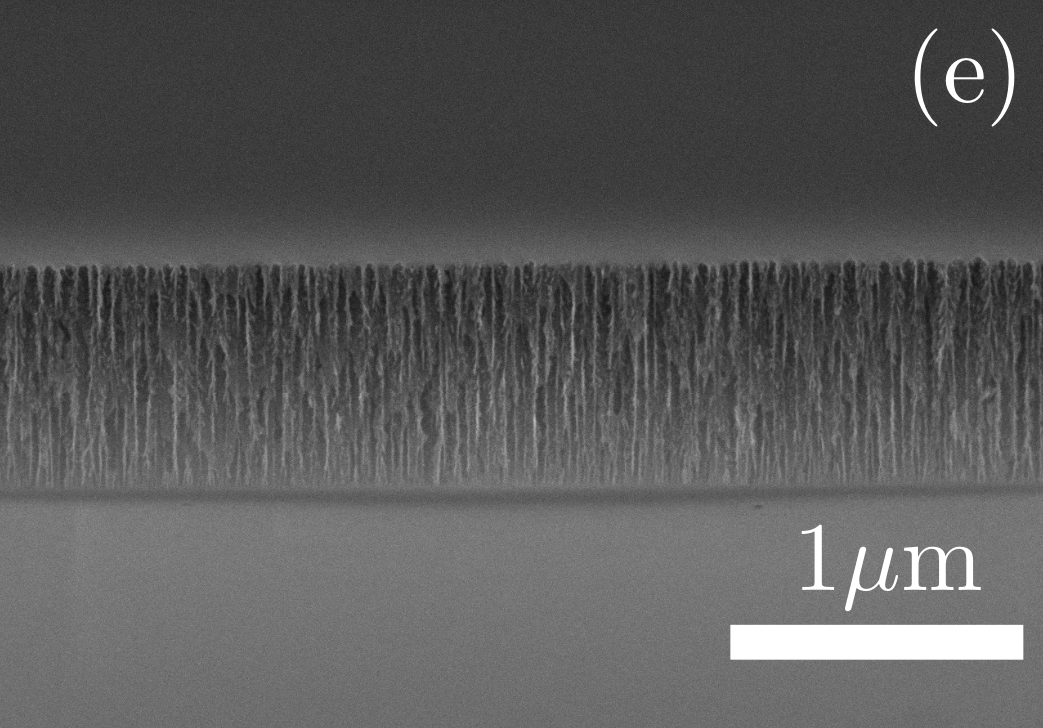}
  \includegraphics[width=.3\textwidth]{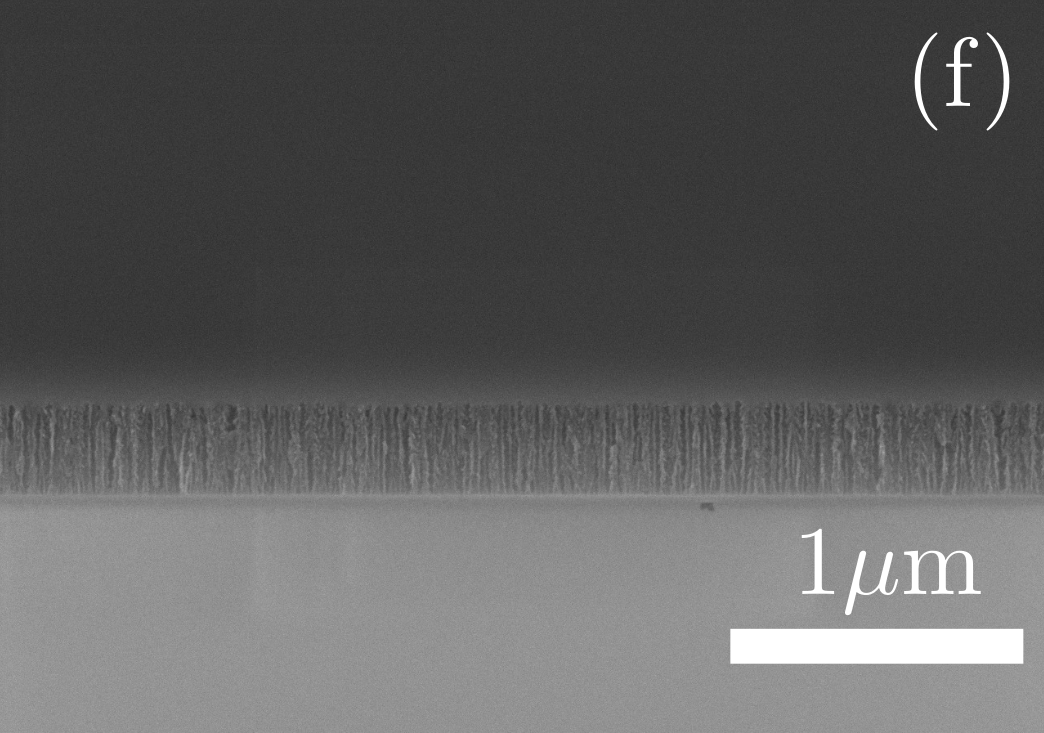}
  \caption{Top ((a), (b), and (c)) and lateral ((d), (e), and (f)) SEM
    micrographs of the sample $S_3$ taken from a region corresponding to the left
    side ((a) and (d)), center ((b) and (e)), and right side ((c) and
    (f)) along the center-line shown in Fig. \ref{fig:SI3}.}
  \label{fig:sem}
\end{figure}

In Fig. \ref{fig:sem} we show some SEM images of the sample $S_3$,
including top views and lateral views taken after
the optical characterization was completed and the sample was cut
along its center-line (see Sec. \ref{sec:R}) from
three regions around the left side, the center, and the right side
of the sample shown in Fig. \ref{fig:SI3}.
We confirm that as we move away from the electrode, the pore
size and pore density diminish (panels (a) to (c) of
Fig. \ref{fig:sem}) and the sample becomes much thinner (panels (d) to
(f)). We remark that these micrographs were taken at three positions
that were not chosen to correspond to the eight locations at which we measured the
optical spectra. Thus, we didn't use SEM measurements for the quantitative
determination of neither the thickness nor the porosity; they were used
for the qualitative verification of the expected (Fig. \ref{fig:DE1})
and obtained (Table \ref{t:jpdv}) results. Actually, one of our
goals is to show that relatively inexpensive reflectance
measurements might be enough to characterize both the porosity and
etching rate variations along the sample.

\subsection{Reflectance}\label{sec:R}

\begin{figure}
  \centering
  \includegraphics[width=\textwidth]{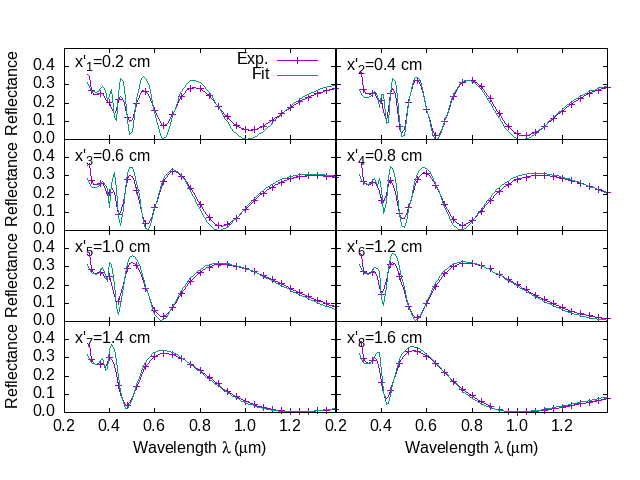}
  \caption{Reflectance spectra at different positions
    ($x'_n=0.2-1.6\,\text{cm}$, $y=0.74\,\text{cm}$)
    on sample $S_1$ ($I_1=5\,\text{mA}$). We show
  experimental results and results fitted through
  equations~\eqref{Eq:ECMR} and \eqref{Eq:Brugg}}.
  \label{fig:R5}
\end{figure}
\begin{figure}
  \centering
  \includegraphics[width=\textwidth]{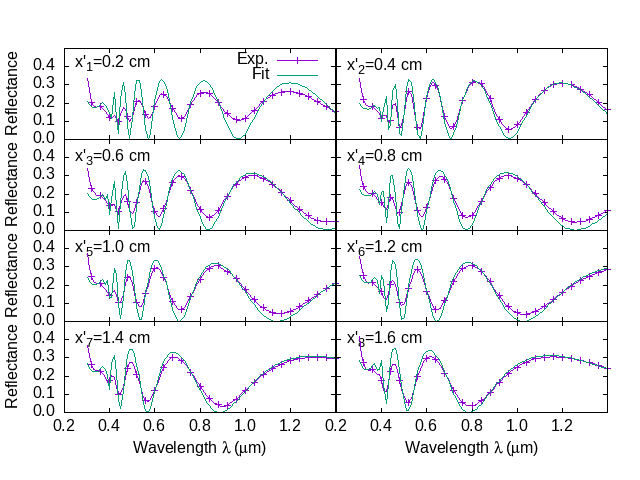}
  \caption{Reflectance spectra at different positions
    ($x'_n=0.2-1.6\,\text{cm}$, $y=0.74\,\text{cm}$)
    on sample $S_2$ ($I_1=10\,\text{mA}$). We show
  experimental results and results fitted through
  equations~\eqref{Eq:ECMR} and \eqref{Eq:Brugg}}.
  \label{fig:R10}
\end{figure}

\begin{figure}
  \centering
  \includegraphics[width=\textwidth]{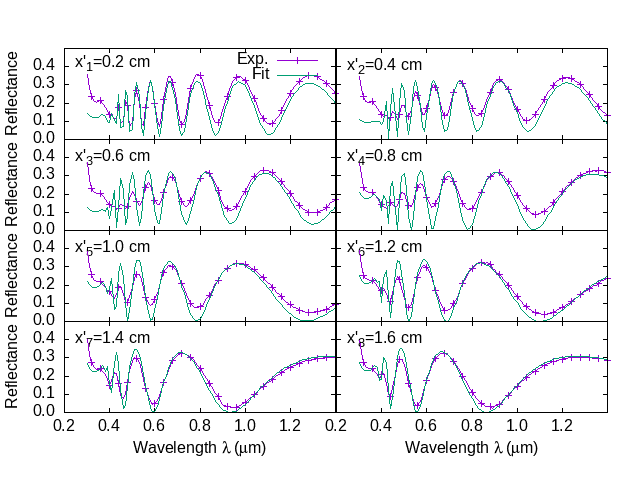}
  \caption{Reflectance spectra at different positions
    ($x'_n=0.2-1.6\,\text{cm}$, $y=0.74\,\text{cm}$)
    on sample $S_3$ ($I_3=20\,\text{mA}$) (Fig. \ref{fig:SI3}),
    We show
    experimental results and results fitted through
    equations~\eqref{Eq:ECMR} and \eqref{Eq:Brugg}}.
  \label{fig:R20}
\end{figure}

We measured the reflectance spectra  at the positions indicated in
Fig. \ref{fig:SI3} on our three samples $S_1$, $S_2$,
and $S_3$, prepared using the same procedure
(Sec. \ref{sec:fabrication-ps-grin}) but applying different currents
$I_1=5\,\text{mA}$, $I_2=10\,\text{mA}$, and $I_3=20\,\text{mA}$. The results are shown in
Figs. \ref{fig:R5}, \ref{fig:R10}, and \ref{fig:R20}.
We fitted the reflectance measurements using Eqs. \eqref{Eq:ECMR} and
\eqref{Eq:Brugg} using the local porosity $p$, and the thickness of the
film  $d$ as
adjustable parameters. To that end we used the {\tt migrad} routine of
the {\tt Minuit} minimization software. From the thickness we
obtained the etching rate $v=d/t$ at each position. Notice that the fits are reasonably
good. The contrast of the reflectance in the experiment is somewhat smaller
than in our fitted curves. We believe this is due to the finite width
$0.7\text{mm}\times2.0\text{mm}$ of the illuminating beam of our
spectrometer. Thus, as our GRIN sample is not macroscopically homogeneous and its
properties have a gradient, the
experimental results have contributions from regions with slightly
different porosities and thicknesses, partially averaging out the interference
maxima and minima. We expect microscopic inhomogeneities such as
interface roughness might also reduce the contrast through scattering
and its corresponding apparent dissipation \cite{missoni2020rough}.
The loss of contrast is most notable for the thicker regions
towards the left side of Fig. \ref{fig:SI3} and for smaller
wavelengths.

The results obtained from the reflectance fits are summarized in table
\ref{t:jpdv}.
\begin{table}
\centering
\renewcommand{\arraystretch}{1.1}
\begin{tabular}{lrrrrr}
  Sample & position & density & porosity &
  thickness & rate\\
  ($I$ (mA)) &$x'_n$ (cm)&  $j$ ($\text{mA}/\text{cm}^2$) & $p$
       & $d$ ($\mu\text{m}$) & $v$ (nm/s)\\
\hline
$S_1$(5) &0.2 &$ 3.61^{+0.01}_{-0.03}   $&$0.56\pm0.03$ &$0.381\pm0.021$ &$  1.52\pm0.08$ \\
         &0.4 &$ 3.36^{+0.09}_{-0.11}   $&$0.59\pm0.02$ &$0.403\pm0.014$ &$  1.61\pm0.06$  \\
         &$0.6$  &$ 2.88^{+0.13}_{-0.14}$&$0.58\pm0.02$ &$0.342\pm0.015$ &$  1.37\pm0.06$ \\
         &$0.8$ &$ 2.32^{+0.14}_{-0.14} $&$0.57\pm0.02$ &$0.278\pm0.011$ &$  1.11\pm0.04$ \\
         &$1.0$ &$ 1.80^{+0.13}_{-0.12} $&$0.56\pm0.02$ &$0.224\pm0.009$ &$  0.90\pm0.03$ \\
         &$1.2$ &$ 1.37^{+0.10}_{-0.09} $&$0.56\pm0.02$ &$0.190\pm0.007$ &$  0.76\pm0.03$ \\
         &$1.4$ &$ 1.06^{+0.07}_{-0.06} $&$0.56\pm0.02$ &$0.153\pm0.005$ &$  0.61\pm0.02$ \\
         &$1.6$ &$ 0.83^{+0.05}_{-0.04} $&$0.56\pm0.02$ &$0.124\pm0.004$ &$  0.50\pm0.02$ \\
$S_2$(10) &$0.2$ &$ 7.22^{+0.03}_{-0.07}$&$0.64\pm0.04$ &$0.686\pm0.054$ &$  2.74\pm0.22$ \\
         &$0.4$ &$ 6.72^{+0.02}_{-0,21} $&$0.64\pm0.02$ &$0.694\pm0.025$ &$  2.78\pm0.10$ \\
         &$0.6$ &$ 5.77^{+0.27}_{-0.28} $&$0.64\pm0.03$ &$0.592\pm0.037$ &$  2.37\pm0.15$ \\
         &$0.8$ &$ 4.64^{+0.28}_{-0.28} $&$0.62\pm0.03$ &$0.530\pm0.034$ &$  2.12\pm0.14$ \\
         &$1.0$ &$ 3.60^{+0.25}_{-0.23} $&$0.61\pm0.03$ &$0.470\pm0.027$ &$  1.88\pm0.11$ \\
         &$1.2$ &$ 2.75^{+0.19}_{-0.18} $&$0.60\pm0.03$ &$0.414\pm0.026$ &$  1.66\pm0.10$ \\
         &$1.4$ &$ 2.11^{+0.14}_{-0.13} $&$0.59\pm0.03$ &$0.350\pm0.018$ &$  1.40\pm0.07$ \\
         &$1.6$ &$ 1.67^{+0.10}_{-0.09} $&$0.58\pm0.02$ &$0.302\pm0.014$ &$  1.21\pm0.05$ \\
$S_3$(20) &$0.2$ &$14.45^{+0.06}_{-0.14}$&$0.69\pm0.03$ &$1.208\pm0.047$ &$  4.83\pm0.19$ \\
         &$0.4$ &$ 13.44^{+0.35}_{-0.41}$&$0.72\pm0.02$ &$1.205\pm0.057$ &$  4.82\pm0.23$ \\
         &$0.6$ &$ 11.54^{+0.53}_{-0.55}$&$0.71\pm0.03$ &$1.050\pm0.058$ &$  4.20\pm0.23$ \\
         &$0.8$ &$  9.29^{+0.57}_{-0.55}$&$0.64\pm0.06$ &$0.766\pm0.088$ &$  3.06\pm0.35$ \\
         &$1.0$ &$  7.20^{+0.49}_{-0.47}$&$0.63\pm0.04$ &$0.540\pm0.037$ &$  2.16\pm0.15$  \\
         &$1.2$ &$  5.50^{+0.38}_{-0.36}$&$0.61\pm0.04$ &$0.449\pm0.035$ &$  1.80\pm0.14$ \\
         &$1.4$ &$  4.22^{+0.28}_{-0.26}$&$0.59\pm0.03$ &$0.369\pm0.018$ &$  1.48\pm0.07$ \\
         &$1.6$ &$  3.33^{+0.19}_{-0.17}$&$0.59\pm0.02$ &$0.338\pm0.014$ &$  1.35\pm0.06$ \\
\end{tabular}
\caption{\label{t:jpdv}
  Current density and fitted parameters for several positions on
  three samples:  porosity, thickness, and etching rate.}
\end{table}
The uncertainties in the fitted parameters $p$, $d$, and $v$ were
chosen as those that increase the sum of squared errors by 10\% with
respect to its minimum value.

\begin{figure}
  \centering
  \includegraphics[width=\textwidth]{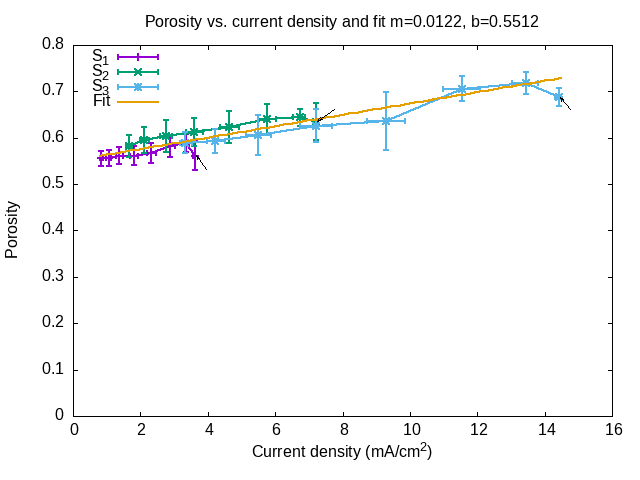}
  \caption{Porosity vs. current density for the three samples
    $S_1$, $S_2$, $S_3$. The arrows point to data that correspond to
    position $x'_1=0.2\,\text{cm}$ (see text). We include a linear fit of the form
    $p=mj_\perp+b$ with parameters
    $m=(0.0122\pm0.001)\frac{\text{cm}^2}{\text{mA}}$ and
    $b=0.551\pm.005$.
    }
  \label{fig:p}
\end{figure}

From the data in Table \ref{t:jpdv} we can plot the fitted porosity vs.
the calculated density current $j_\perp$, as shown for our three
samples in Fig. \ref{fig:p}.
Notice that for each sample the data points lie around straight lines,
except for one outlier point in each corresponding to position
$x'_1$. As discussed in Sec. \ref{sec:ps-grin-single}, those points
were very close to a wall of the electrolytic cell, so they were
affected by leakage currents towards the underside of the wall.
For this reason, we exclude those points from the following
analysis. We made a linear fit of the form $p=mj_\perp+b$ obtaining
\begin{equation}\label{eq:p}
  p\approx (0.0122\pm0.001)\frac{\text{cm}^2}{\text{mA}} j_\perp+(0.551\pm.005).
\end{equation}
Though the fit is reasonably good, it is noticeable that most of the data from
sample $S_2$ lie to the left of the fitted line while most of the data
from the other samples lie to the right. This discrepancies may be due
to the uncertainty in the positions $x_n$ (see Table \ref{t:xj}) due
to the ill defined distance from the edge of the sample (which roughly
corresponds to the position of the o-ring) to the edge of the cell. A
deformation of the o-ring when the cell is pressed against the wafer
would be equivalent to a rigid shift of the sample, i.e., to adding a
constant to all the positions $x_n$. Thus it is likely that the sign
and magnitude of that uncontrolled constant displacement changes among
samples.

\begin{figure}
  \centering
  \includegraphics[width=\textwidth]{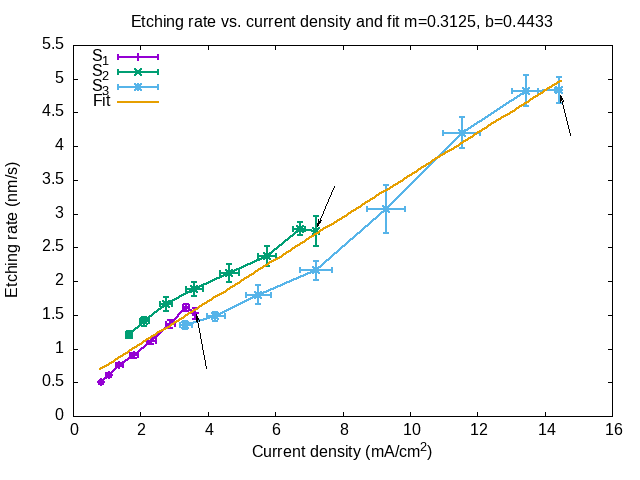}
  \caption{Etching rate vs. current density for three samples
    $S_1$, $S_2$, $S_3$. The arrows point to data that correspond to
    position $x'_1=0.2\,\text{cm}$ (see text). We include a linear fit of the form
    $v=mj_\perp+b$ to the data with parameters $m=(0.312\pm
    0.017)(\text{cm}^2/\text{mA})(\text{nm/s})$ and
    $b=(0.44\pm0.09)\text{nm/s}$. }
  \label{fig:v}
\end{figure}

Similarly, we can plot the fitted etching rate vs.
the calculated density current, as shown for our three
samples in Fig. \ref{fig:v}.
As for the porosity, excluding the points in the
immediate proximity of the wall of the cell, the remaining points
lie roughly on straight lines for each sample, and they
may all be approximated by a linear fit to all the data, from which we get
\begin{equation}\label{eq:v}
  v\approx (0.312\pm0.017)\frac{\text{cm}^2}{\text{mA}}
  \frac{\text{nm}}{\text{s}}j_\perp +
  (0.44\pm0.09)\frac{\text{nm}}{{s}}.
\end{equation}
Similarly to
Fig. \ref{fig:p}, most data from sample $S_2$ lies to the left of the
fitted line and most of the data from $S_1$ and $S_3$  lie to the
right, which could be similarly explained and corrected.
These corrections are different for layers grown on different samples,
but for a multilayered system, produced by modulating the current $I$ in
time, a single correction ought to be enough for all the layers.

\begin{figure}
  \centering
  \includegraphics[width=\textwidth]{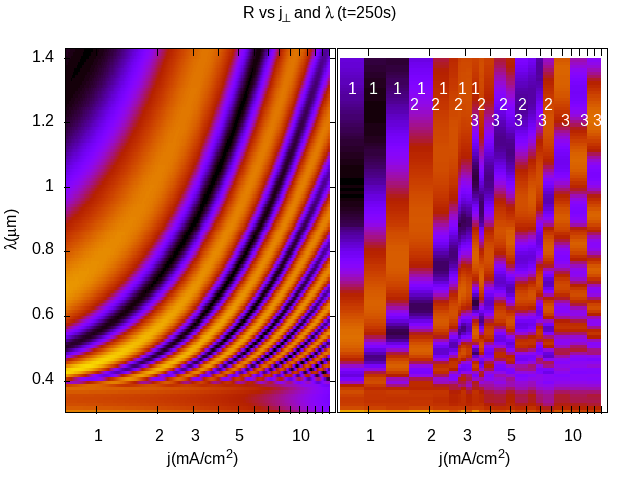}
  \caption{Reflection $R$ of a PS film as a function of the current density
    current $j_\perp$ and wavelength $\lambda$ for an etching time
    $t=250\,\text{s}$. Theory (left panel) and experiment (right
    panel, see text). We indicate with the numbers 1, 2, 3 the
    data corresponding to the corresponding samples $S_1$, $S_2$
    (shifted, see text) and  $S_3$.}
  \label{fig:Rvsjl}
\end{figure}

Using Eqs.  \eqref{Eq:ECMR}-\eqref{eq:v} we can calculate the
reflectance spectra for any
desired current density $j_\perp$ and etching time. This is
illustrated in Fig. \ref{fig:Rvsjl} for the parameters that correspond
to our samples.
On the right hand side of the figure we display a quilt made by
patching together the experimental reflectance data corresponding to
Figs. \ref{fig:R5}- \ref{fig:R20}, ordered according to the
corresponding current density as presented in Table
\ref{t:jpdv}. Nevertheless, for the reasons mentioned above, we
eliminated the data corresponding to position $x'_1$ on all
samples. Furthermore, as discussed above and for visualization
purposes we applied small rigid translations to the samples along the
$x$ direction, thereby modifying $j_\perp$, before patching together
the reflectance spectra.
Notice that the experimental results may
be grouped along bright and dark bands that correspond closely to
those expected theoretically, though the experimental bands are somewhat shifted,
especially for low current densities, and their minima are not as low.

The theoretical reflectance could as well be calculated using Eqs.
\eqref{Eq:Brugg}-\eqref{eq:v} and a transfer matrix approach for more
complicated structures, in which the
current $I$ is modulated in time, allowing the design and analysis of
non-trivial multilayered GRIN PS structures.

\section{Conclusions}
\label{sec:conclusions}
By choosing a simple geometry for our electrochemical cell, namely, a
rectangular prism, and employing a point-like electrode, we were able
to derive an expression for the calculation of the electric current
density along the surface of a Si sample. The expression was obtained
by using image charge theory and a plane-wise summation of Coulomb
potentials in 2D reciprocal space, and it is made up of
a rapidly converging sum of analytical terms, just a few of which
need be kept. This allows the characterization
of the porosity of a PS layer and the etching
rate for different values of the etching current density by making
optical measurements on several positions of just a
few samples. We illustrated the procedure with three samples on each of which we
measured reflectance spectra at eight different positions.
Through an optimization procedure, we fitted the porosity and etching
rates corresponding to each current density. We used these data to
obtain calibration curves that allow the approximate prediction of the
properties of porous Si samples prepared with a current density that
spans a range of more than one order of magnitude. The
small discrepancies we found between our samples and the fit
are probably due to our unfortunate choice of reference point for
measuring positions, as we used as reference the ill defined edge of
the sample, while the important reference for our calculation is the
edge of the wall of the electrolytic cell. Nevertheless, these
discrepancies are reduced through a rigid shift of the coordinates
$x_n$ used in the analysis. Our calibration curves can be used to
design and calculate the optical
properties of other systems prepared under similar conditions, such as
multilayered GRIN systems from which photonic crystals, micro-cavities
and sensors with position-dependent properties. This is the subject of
ongoing work. As the etching rate for the formation of porous silicon
and the resulting porosity are very sensitive to the growth conditions
and to the properties of the substrate, such as its doping,
there are no universal calibration curves. Thus, it is useful to have
a procedure that from just one or a few samples can produce calibration curves
that may be employed for other samples grown under similar conditions.

\section*{Acknowledgments}
\label{sec:acknowledgments}
CAOD is grateful for a scholarship from CONACyT and to
V. Castillo-Gallardo and L.E. Puente Díaz for useful discussions. VA acknowledges support
from CONACyT Basic Sciences project A-S1 30393 and from PRODEP, and is
grateful to ICF where she spent a sabbatical.
WLM acknowledges the support of DGAPA-UNAM under grant
IN111119.

\printcredits

\bibliographystyle{unsrt}
\bibliography{bibliography}
\end{document}